\documentstyle[12pt]{article}  
\def\etal{{\it et al} \ }
\def\kms{km s$^{-1}$}
\def\kmsmpc{km s$^{-1}$ Mpc$^{-1}$\ }
\def\hmpc{{\rm h}$^{-1}$ {\rm Mpc}\ }
\def\Mo{{\rm M_\odot}}

\def\gsim{ \lower .75ex \hbox{$\sim$} \llap{\raise .27ex \hbox{$>$}} }
\def\lsim{ \lower .75ex \hbox{$\sim$} \llap{\raise .27ex \hbox{$<$}} }
\def\pp{\noindent\parshape 2 0truecm 16.0truecm 1.0truecm 15truecm}

\def\spose#1{\hbox to 0pt{#1\hss}}
\def\simlt{\mathrel{\spose{\lower 3pt\hbox{$\mathchar"218$}}
     \raise 2.0pt\hbox{$\mathchar"13C$}}}
\def\simgt{\mathrel{\spose{\lower 3pt\hbox{$\mathchar"218$}}
' \raise 2.0pt\hbox{$\mathchar"13E$}}}
\def\oneskip{\vskip\the\baselineskip}

\begin{document}
\begin{titlepage}
\baselineskip 15pt
\centerline{\LARGE The Local Group as a test of cosmological models}
\oneskip
\oneskip
\centerline{
Fabio Governato$^{\bf 1}$, Ben Moore$^{\bf 2}$, Renyue Cen$^{\bf
1,3}$, Joachim Stadel$^{\bf 1}$, George Lake$^{\bf 1}$ \& Thomas
Quinn$^{\bf 1}$}
\oneskip

${\bf ^1}$ Department of Astronomy, University of Washington, Seattle,
USA

${\bf ^2}$ Department of Physics, University of Durham, South Road,
Durham, UK

${\bf ^3}$ Princeton University Observatory, Princeton, USA

\vskip 3cm

\begin{abstract}
\baselineskip 22pt

{\bf

The dynamics of Local Group and its environment provide a unique
challenge to cosmological models. The velocity field within 5$h^{-1}$
Mpc of the Local Group (LG) is extremely ``cold''. The deviation from
a pure Hubble flow, characterized by the observed radial peculiar
velocity dispersion, is measured to be $\sim$ 60\kms.  We compare the
local velocity field with similarly defined regions extracted from
N-body simulations of Universes dominated by cold dark matter (CDM). This
test is able to strongly discriminate between models that have different
mean mass densities.  We find that neither the $\Omega=1$ (SCDM) nor
$\Omega=0.3$ (OCDM) cold dark matter models can produce a single
candidate Local Group that is embedded in a region with such small
peculiar velocities.

 For these models, we measure velocity dispersions between
300\---700\kms\ and 150\---300\kms\ respectively, more than twice the
observed value.

Although both CDM models fail to produce environments similar to those
of our Local Group on a scale of a few Mpc, they can give rise to many
binary systems that have similar {\it orbital} properties as the Milky
Way--Andromeda system.

The local, gravitationally induced bias of halos in the CDM ``Local
Group'' environment, if defined within a sphere of 10 Mpc around each
Local Group is $\sim$ 1.5, independent of $\Omega$. No biasing scheme
could reconcile the measured velocity dispersions around Local Groups
with the observed one. 

Identification of binary systems using a halo finder (named Skid
 (hhtp ref: http://www-hpcc.astro.washington.edu/tools/DENMAX))
 based on a local density maxima search instead of a
 simple linking algorithm, gives  a much more complete sample.
We show that a standard ``friend of friends'' algorithm would miss about 
40\% of the LG candidates present in the simulations. }

\end{abstract}

\end{titlepage}

\section{Introduction}

Galaxies are observed to lie in a wide range of environments, from
rich virialized galaxy clusters, to small groups that contain only two
galaxies.  The most ``typical'' galactic environment is a group that
has a total luminosity of about $6L_*$, where $L_*$ is the
characteristic break in the galaxy luminosity function (Moore, Frenk
\& White 1993).  (For H$_0$ = 50\kmsmpc,
$L_*\approx4\times10^{10}L_\odot$ in the B band) The LG is a very
typical small galaxy group, dominated by the Milky Way \-- Andromeda
binary system and with over two dozen smaller satellites; the total
luminosity is about $2.5L_*$.

Although the LG is not virialized and appears to reside in mildly
overdense region of the
Universe, the nearby larger Virgo cluster and the presence of the
filamentary local supercluster, which itself connects to the Great Attractor
( e.g., IRAS density field; Strauss \etal 1992), creates a complex
 gravitational tidal field.

Measuring galaxy masses and $\Omega$ using Local Group dynamics has a
rich history. Applying the timing argument to the Milky Way and
Andromeda (Kahn \& Woltjer 1959), a lower limit on mass of order
$3\times 10^{12}\Mo$ is recovered using the radial infall
solution. The retardation of nearby galaxy velocities from a pure
Hubble expansion was used by Giraud (1986) and Sandage (1986) who both
concluded that the mass of the Local Group must not be larger than a
few times $10^{12}M_\odot$. 
On a larger scale, the LG motion towards
the Virgo cluster, has been estimated to be between
 85km/s (Faber \& Burstein 1988,  modeling the
local velocity field with the Great Attractor) to 350km/s ( with an average
infall velocity of $250\pm 50$km/s (the error is indicative of both
the differences in different analyses and the internal measurement
errors; Yahil \etal 1980; Aaronson \etal 1982; Dressler 1984;
Kraan-Korteweg 1985; Huchra 1988). 
This value  has been used to constrain $\Omega$,
on a scale of 10\hmpc, yielding a fairly large acceptable range of
$\Omega$ from 0.2 to 1.0, due to large uncertainties on both the
observed Virgocentric infall motion itself and the mass density
distribution in the region (Cen 1994).
Peebles (1989a) calculates the
orbits of the nearby galaxies using the ``least action principle",
assuming that peculiar velocities at high redshifts are negligible,
and then integrating orbits forwards to try and reproduce their
present day positions and velocities. Using only the LG members, both
high and low $\Omega$ solutions are equally likely, but when galaxies
out to 4000\kms\ are used a low value of $\Omega$ is preferred
(Peebles 1995; Shaya \etal 1995).
Numerical simulations have played an important role in evaluating and
testing these methods. Similar environments can be extracted from
large N-body simulations and the same techniques can be applied to the
``observed'' information and the results compared with the true
properties (Bushouse \etal 1985; Moore \& Frenk 1990; Kroeker \&
Carlberg 1991; Schlegel, Davis \& Summers 1994, SDS thereafter; Cen 1994;
 Branchini \& Carlberg
1994).  Typically, errors in the mass estimators are at least a factor
of 2. The uncertainties result from a lack of the full velocity
information, complex gravitational fields generated by surrounding
structure and the assumption that the mass in halos has always been
constant with time (Dunn \& Laflamme 1995).

The interplay between fluctuations on different mass scales can be
captured using large-scale N-body simulations hence the latter are
ideally suited for studies of this nature.  Because of its proximity,
we know a great deal more about our Local Group than for any other
galaxy group, e.g., the true distances and peculiar velocities are
known for many of the nearby galaxies. Peebles has often emphasized
using the Local Group to constrain cosmological models, and noted that
both the galaxy distribution and the local velocity field may provide
a challenge to current models (Peebles 1989b).

The ``coldness'' of the local Hubble flow can be quantified by
measuring the dispersion (or rms) of the observed peculiar velocities
around the mean local Hubble flow.  Using 31 galaxies outside of the
Local Group but within 5$h^{-1}$Mpc, SDS (1994) find a value
 of 84 \kms\ (60\kms\ when distance errors
are kept into account).  A similar value was also obtained by Giraud
(1986).  This remarkably cold velocity field is only a local effect,
and we do not know if it is a property of ``Local Group'' environments
in general.  Averaged over larger volumes of the Universe, the mean
pairwise dispersion of galaxies increases to a few hundred \kms on
the same scales if IRAS galaxies are used (Fisher {\it et al} 1994)
to almost 700 \kms if a sample including more galaxies in clusters is
used (Marzke {\it et al} 1995, Guzzo {\it et al} 1996, Somerville, Davis 
and Primack 1996).
The fact that there are no nearby blue-shifted galaxies within a
sphere of 5h$^{-1}$ Mpc is another indicator of a fairly uniform
Hubble flow. 
SDS searched for Local Group candidates in N-body
simulations of CDM and mixed dark matter (MDM)
Universes and concluded that the MDM model
 provides more candidates because the peculiar
velocities are suppressed by the free streaming of the hot
component. However, two major drawbacks of that work are the small volumes that
were simulated and rather than defining Local Groups as binary systems,
single galaxies in low density regions were selected instead.

In this paper we use high resolution large-scale N-body simulations,
that resolve single $L_*$ galaxy sized halos with hundreds of
particles, while sampling a volume sufficiently large to contain
hundreds of such halos.  This allows us to construct large samples of
``Local Groups'' identified in a more robust and realistic way, such
that they are a better match to the local dynamics and environments of
our Local Group. We identify binary groups starting from a halo
catalog built using a new halo-finding algorithm that eliminates the
intrinsic biases produced by friend-of-friends algorithms. The
peculiar velocity fields surrounding these mock LGs extracted
from both high and low $\Omega$ CDM Universes are then compared with
the real observed data.  Our aim is to determine if a Universe
dominated by CDM can produce candidate sites that could
host the Local Group galaxies, as well as matching the local velocity
field. We shall also use these candidate LGs to test the
``Virgo-centric infall'' technique for measuring $\Omega$.

\section{The N-body simulations}

Our simulations were performed using PKDGRAV (Dikaiakos \& Stadel 1995,
Stadel \& Quinn 1997), a
portable parallel treecode that supports true periodic boundary
conditions,  required for  simulations of cosmological
volumes.  Although the simulations here would be considered fairly
large by current single processor standards, they represent relatively
small investments of ``wall clock'' time ($\sim140$ hours), using
PKDGRAV on 64 processors of a CRAY T3D parallel computer.

We ran two simulations: a CDM $\Omega$ = 1, h = 0.5, $\sigma_8$ = 0.7
model (SCDM) and a CDM $\Omega=0.3$, h = 0.75, $\sigma_8$ = 1 model
(OCDM) with zero cosmological constant.  Each simulation cube is 100
Mpc on a side with both using 144$^3$ particles and a spline kernel
softening of 60 kpc (The spline kernel is completely Newtonian at 2
softening lengths).  Since our goal is to identify galactic sized dark
halos, not to greatly resolve their internal structure, this force
resolution is adequate.  We used 650 and 1000 timesteps in SCDM and
OCDM simulations, respectively.  The mass resolution is such that a
well resolved  halo (30 particles)  has a circular
velocity of 110 and 90 \kms  in the two models. We can
therefore resolve halos that are associated with galaxies that have
luminosities as low as 5\% of an $L_*$ galaxy.

The adopted SCDM and OCDM models are known to be ``wrong", {\it i.e.}
they cannot satisfy all of the observational constraints over a wide
range of scales (for a review see Ostriker 1993).  However, they are
sufficiently realistic and in many aspects match observations very
well so that they have been used frequently as convenient testbeds for
``benchmarking" other simulations or for comparative studies.  The
normalizations we have used for both models are roughly correct in
terms of reproducing the correct abundance of rich galaxy clusters
(Bahcall \& Cen 1992; Oukbir \& Blanchard 1992;
Viana \& Liddle 1995; Kochanek 1995; Eke, Cole \& Frenk 1996; Bond \&
Myers 1996), although the amplitude of the SCDM model may be slightly
too high.  Note that, while the OCDM model is approximately
COBE-normalized (Gorski \etal 1995), the value of $\sigma_8$ for a
COBE-normalized SCDM model would be twice  what we use here (Bunn,
Scott, \& White 1995).  However, a slight tilt ($0.1-0.2$) from the
Harrison-Zeldovich spectrum ($n=1$) would suffice to make the SCDM in
accord with both COBE and galaxy cluster observations.  In any case,
our conclusion would not sensitively depend on $n$ and our two adopted
models should give a good representation of the Universe on small to
intermediate scales.

\section{Finding halos and local group candidates}

The local overdensity can be determined using the complete catalogues
of IRAS galaxies. Although these galaxies avoid the centers of rich
clusters, they trace the rest of the galaxy distribution very well
(Strauss \etal 1992).  For the 1.2Jy survey, SDS calculate
that $\delta \rho /\rho \sim 0.25$ for a 5h$^{-1}$ Mpc sphere centered
on the Local Group, normalized to the mean IRAS galaxy density.  The
current best estimate based on IRAS galaxies is $\delta \rho
/\rho=0.60\pm 0.15$ for a top-hat sphere of radius $500$km/s (Strauss
1996, private communication).  Hudson (1993) uses a compilation of
optical galaxy surveys to study the local density field and within the
same volume he finds an overdensity of $\sim 0.2$.  The agreement
between the optical and IRAS surveys is encouraging.

Our simulations only follow the evolution of the dark matter
component. It is expected that baryons condense and form galaxies at
the centers of the dark matter halos, a premise that is supported by
simple considerations of the relevant dynamical timescales or by
numerical simulations that include a gaseous component (White \& Rees
1978; Cen \& Ostriker 1992,1993a,b; Katz, Hernquist \& Weinberg 1992).
Therefore for our purpose of locating LG halo candidates (not
resolving their internal structure), what we need is an algorithm that
can find dark halos within the simulation by grouping together
particles in an appropriate manner.  Throughout this paper we assume
that a galaxy with an observed circular velocity, $V_c$, would be
found in a dark halo that has a similar circular velocity.

Linking together all particles within a minimum distance using the
classic friends-of-friends type algorithm (FoF) has the undesirable
consequence of linking together binary pairs of halos. This pathology
is avoided by using a grouping algorithm that ``moves particles"
towards local density maxima to identify halo membership (Gelb \& Bertschinger
 1994a; see Fig.1 \& 2).  A complete description of the algorithm, named
Skid (hhtp ref: http://www-hpcc.astro.washington.edu/tools/DENMAX), is
given in Stadel \etal (1996).  Previous analyses of binary halos must
be regarded with caution since we find that almost twice as many
binary systems are uncovered using the new algorithm; catalogues of
binary halos created with friends-of-friends (http ref:
http://www-hpcc.astro.washington.edu/tools/FOF) may be biased, for
example, due to exclusion of close pairs.

The Skid algorithm breaks up clusters and finds individual halos
orbiting within larger halos.  In order to calculate the total mass of
a larger halo, (for example to get the mass within the collapsed region, 
 neglecting any substructure) 
 we must re--group the particle distribution using a FoF
algorithm.  This is particularly important when we are trying to
determine the presence and mass of Virgo sized clusters that lie near
candidate Local Groups.

The FoF linking length was set to select particles within
overdensities $>125$ with respect to the critical value, which is the
overdensity at the virial radius calculated using the spherical infall
model (Gunn \& Gott 1972).  The density field for Skid was obtained by
smoothing the particle distribution with a spline kernel over the
nearest 32 neighbours for each particle. All particles with a local
density larger then 57 times the critical density (the local density at the
virial radius in an isothermal sphere)
were then regrouped into halos. To avoid spuriously
small halos due to the granularity of the particle distribution, halos
within 3 times the softening length of each other were linked
together, which sets the minimum separation (180 Kpc) between galaxy
members in our binary samples. This is not a problem as we are interested
 mainly in binaries with separations larger than that.

No restrictions were placed on the local environment of the candidates,
and we created three separate catalogues of binaries for each
cosmological model with the following three different (increasingly
stringent) constraints:

\noindent 1) A generic sample of binary halos with separations $s <
1.5$ Mpc and circular velocities $125 < V_c/{\rm km s}^{-1} < 270$ (cat1).

\noindent 2) A LG sample defined such that $s < 1.0$ Mpc,
circular velocities $125 < V_c/{\rm km s}^{-1} < 270$, negative radial
velocities and no neighbors within 3 Mpc with circular velocity larger
than either of the two group members (cat2).

\noindent 3) A LG sample defined as (2) but 
with the additional requirement that the binary halos must lie close 
(5-12h$^{-1}$Mpc) to a Virgo sized cluster
 with $500 < V_c/{\rm km s}^{-1} < 1500$ (cat3).

\vskip 0.2in

\centerline{\bf TABLE 1}
\smallskip
\centerline { LG abundance }
\bigskip
\vbox{\hbox to \hsize{\hfil\vbox{\halign {#\hfil&&\quad\hfil#\hfil\cr
\noalign{\hrule}\cr
\noalign{\smallskip}
\noalign{\hrule}\cr
\noalign{\medskip}\cr
Model & Skid & FoF \cr
\noalign{\smallskip}
\noalign{\hrule}\cr
\noalign{\medskip}\cr
SCDM \hfil  cat1  & \hfil 501 & \hfil 311 \cr
OCDM \hfil cat1  & \hfil 211 & \hfil n/a \cr
SCDM \hfil cat2  & \hfil 59 & \hfil 35 \cr
OCDM \hfil cat2  & \hfil 17 & \hfil n/a \cr
SCDM \hfil cat3  & \hfil 6 & \hfil n/a \cr
OCDM \hfil cat3  & \hfil 2 & \hfil n/a \cr
\noalign{\medskip}\cr
%\noalign{\hrule}\cr
\noalign{\medskip}}}\hfil}}
\medskip

The FoF algorithm finds 40\% less binary groups than Skid, a fact that
reflects the pathology of the latter in linking close pairs of halos
together.  The OCDM model creates more than a factor of two fewer
Local Group candidates than SCDM.  The reason for this difference
owes primarily  to the different $\Omega$ in the two models,
which  is partially balanced by the adopted values of $\sigma_8$.

The lower $\Omega$ model contains fewer objects at a mass scale of
$\gsim 10^{13}M_\odot$, {\it i.e.} the mass of the Local Group.  This
number density difference at this particular mass scale and at the
present time is consistent (Governato {\it et al.} 1996) with an
estimate made using the Press \& Schechter theory (Lacey \& Cole
1993).

In both models, the number density of binary halos increases with
redshift, doubling with respect to the present at z $\sim 1$.  This
is partially due to an increasing number of collapsing objects at the
LG mass scale at earlier times and to the fact that it is easier for
the LG binary systems to pass the isolation criteria from more massive
groups.

\section{Results}

Is the Local Group a unique system in the Universe?  If we occupied a
special place then a measurement of physical quantities in our
neighborhood would be less meaningful.  Small galaxy groups are the
typical environment that galaxies inhabit and Karachentsev (1996)
argues that the Local Group is similar to other nearby groups of
galaxies.  A key test is the ability of hierarchical clustering models
to produce structures similar to the LG at the present time, both in
terms of producing binary systems similar to the Milky Way--Andromeda
systems and in terms of reproducing the characteristics of the
surrounding environment of the LG.

We tested this by first selecting a generic binary sample 
 with constraints looser than those we used to define the LG binary
systems (using criterium (1)). 
The idea is to create a ``control sample'' and check first if
our subsamples of LG binaries have extreme, or special dynamical
properties with respect to this more general sample.

\subsection {The dynamics of the binary systems}

There are no discernible differences between the dynamical properties
of the generic sample of binary halos and the sample of mock ``Local
Groups". The LG sample shows the same circular velocity  distribution 
( which is directly related to the mass distribution)  of the generic
binary sample (Fig. 3). The distribution of radial velocities is consistent
with the larger distribution of the generic sample (Fig. 4).  Also,
the observed radial velocity of the Andromeda-Milky Way system
(-120\kms) is in very good agreement with our LG sample that only has
the constraint of {\it negative} radial velocity.  

Fig. 5 shows the
relative radial peculiar velocities plotted against the separation of
the halos.
There is a weak trend in that closer pairs have larger negative
velocities, and the SCDM model has a larger scatter than OCDM.
Furthermore, there are many more systems in the SCDM model where the
relative radial peculiar velocities are very large. We believe that
this is a result of the stronger tidal field present in the SCDM
model.  A visual inspection of the most extreme cases reveals that
pairs with very high negative relative velocities (larger than 400
\kms) have massive nearby companions, often just outside the exclusion
zone (3 Mpc). For these systems, applying the ``timing argument'' will
clearly yield false mass estimates.

As a measure of the ellipticity of the LG orbits, in Figure 6 we plot
a histogram of the radial velocities divided by the absolute value of
the relative (vector) velocity.  Again, we observe no difference
between SCDM and OCDM. However, the candidate LG halos tend to have
radially biased orbits. This is probably due to the isolation criteria
so that the motions are not strongly perturbed by nearby massive
halos. However in all samples there is a wide spread in orbital
eccentricities, with some pairs on fairly circular orbits.  So, while
these results support the picture of a Milky Way-Andromeda system
being more likely on a radial orbit and at its first pericentric
passage, orbits with smaller eccentricities cannot be excluded, at
least on statistical grounds.

Producing binary system with {\it dynamical} characteristics similar
to the LG is not a problem for CDM models. This reflects a generic
success for CDM--like hierarchical structure formation models that
produce large numbers of late collapsing objects at a scale of
$\sim$10$^{13}\Mo$.  We shall show that this result is not true once
we take into account the properties of the local environment of the
model Local Group candidates.

{\subsection {The environment of Local Groups}}

The topology of the environments of the simulated Local Groups can be
visualized via a series of MPEG movies that accompany this paper. In
general, Local Groups prefer to inhabit low density regions, sometimes
with an overdensity that is below average, or on the outskirts of
clusters, but never in very dense environments (Fig. 7 \& 8 and
Movie 1).  Also visible on a visual inspection is a tendency for the
candidate Local Groups to reside in filaments and planes (Fig. 9,
Movie 2).  It would be interesting to quantify this statement by
applying a statistic such as that  used by Brandenberger \etal (1993)
to candidate systems in higher resolution simulations.  Fig. 10 plots
the histogram of local densities around each LG binary measured in a
sphere of 10Mpc (h already accounted for). 
 The OCDM distribution seems to be marginally more
peaked around values close to the average density, while that of the
SCDM is  broader.

For each candidate Local Group we plot the overdensity within 5$h^{-1}$ 
Mpc against the dispersion in peculiar velocities of all the
{\it halos} within this region (Fig.11). Also plotted is the observed
value for our Local Group.  Each individual value was obtained fitting
locally the value of the Hubble constant (starting from the smallest
galaxy in the LG pair) and then subtracting the peculiar velocity of
every halo previously identified.  A 5$h^{-1}$ Mpc sphere corresponds
to different regions in velocity space for SCDM and OCDM because the
two models have different values for the Hubble constant. However, we
found no difference in our results when we used a sphere of 10 Mpc (no
$h$ dependence) centered on the LG's within the OCDM model.  A more
sophisticated approach might use a more detailed model of the Hubble
flow close (within $\sim$ 2h$^{-1}$Mpc) to each LG binary. However, we
have checked that the results do not change if galaxies within
2.5h$^{-1}$ Mpc from each LG are discarded, where this correction could
be of some importance.

The OCDM candidates lie in a well defined region of this diagram, and
all have higher peculiar velocity dispersions than measured. The SCDM
groups have more scatter, and almost all of the points lie above those
of OCDM.  The same relation holds if the halo density is used instead
of the mass density, implying that there is a tight (even if biased,
 see $4.4$) relation between the halos and the mass distribution. 
Typically each sphere contains several tens of halos.

The observed dispersions were obtained after merging together and
taking the average value for all galaxies within groups. This is
similar to measuring the dispersion in the Hubble flow using DM halos,
as is done in this paper, because substructure within groups sized halos
is erased due to overmerging (Moore et al 1996).

Our main result is that for SCDM, we find that for Local Group
candidates at an overdensity of 0.2, the peculiar velocities of nearby
halos have a dispersion between 300 \--- 700 \kms. OCDM candidates
give candidates with dispersions in the range 150 \--- 300\kms,
several times larger than the observed value. Neither of these models
are able to produce a single LG with a local velocity dispersion
comparable with the observed value. There is a trend for LGs living in
underdense regions to be surrounded by a colder Hubble flow, however
even in such regions the flow is hotter than the observed one.  This
places a strong constraint on both models, since even invoking a
generous bias that would stretch the distribution of points in Fig. 11
 along the $x$ axis, would not lower the values of the measured
velocity dispersions. If we use the new suggested value for the local 
overdensity  (i.e. 0.6, as mentioned in paragraph 3)
 then the comparison with real data would be even worse for both CDM models.

Lower values of $\Omega$ might close the gap between the peculiar
velocity dispersion surrounding Local Group candidates in the models
and the observed value. Indeed, low values of $\Omega$ are inferred
when the dynamics of galaxy clusters are considered. However, 
measurements of large scale streaming velocities indicate higher 
values, $\Omega\gsim0.3$ (Nusser \& Dekel 1993 ). 
Also, recent numerical studies of the cluster velocity field based
on LG--like observers (Moscardini {\it et al.} 1996)  favor
$\Omega \gsim 0.4$ in CDM models. 

{\subsection {Virgo-centric infall}}

The component of the Local Group's velocity towards the Virgo cluster
can be used to infer $\Omega$ on a scale $\sim 10$ Mpc (see e.g.,
Huchra 1988 for a review).  We define Virgo sized clusters as those 
halos identified by FoF
with 1-d velocity dispersion in the range 550-750\kms, which yields 4
such ``Virgos'' in each of our volumes. As is also discussed below,
 FoF is  better suited to identify clusters with a given mass.
In this case  we did not want to break each  ``cluster'', i.e. each region
 within the overdensity of 57 in individual subclumps, as Skid would have done.

We apply a modified relation
of the linear theory spherical infall model (Gunn 1978) including
nonlinear effects (Yahil 1985, see also Villumsen \& Davis 1986) to
our model Local Groups that lie within 1000 \kms\ of a Virgo sized
cluster:
\begin{equation}
{\Delta H r\over H r} =-{f(\Omega)\over 3}{\Delta\rho\over
\rho}(r)[1+{\Delta\rho\over\rho}(r)]^{-0.25},
\end{equation}
\noindent where
$H r$ is the flow at radius $r$ from the center of the Virgo cluster,
$\Delta H r$ the deviation from the mean Hubble flow,
$f(\Omega)\approx \Omega^{0.6}$, $\Delta \rho/\rho(r)$ the mean
overdensity inside the Virgocentric radius $r$.  The last term in the
right hand side of equation (1) reflects a fit to the nonlinear
effects.
The six LG candidates in the SCDM (Fig.12, panel (b))
 simulations give a range of
inferred mass estimates that differ from the true mass by at least a
factor of 2, although the mean value is roughly correct.  As OCDM has
a larger Hubble constant, the physical volume around Virgos that is
useful for estimating the Virgo infall is smaller and so we
have only two candidates in this case. These would yield $\Omega=0.05$
and 0.29 for observers in their Local Group vantage points (Fig.12, panel (a)).
Similar conclusions were obtained by Bushouse \etal 1985, Cen 1994.

Inaccuracies in the mass estimates using spherical infall theory arise
from several effects.  Infall velocities are not perfectly radial
since the gravitational field on 10$h^{-1}$ Mpc scales is very complex.
Defining the center and peculiar velocity of a cluster is difficult
since large pieces of substructure may be moving rapidly within a
clusters halo. Skid will identify a largest subclump that may have a
peculiar velocity with respect to the whole cluster between 500 \---
1000\kms. The same thing may happen in the real case, as the
 Virgo cluster is itself a complex system with substructure
and smaller groups that appear to be extended along the
line of sight and can be broken up into several merging sub-groups
(Bohringer {\it et al.} 1994). Therefore, we cannot expect
 to infer $\Omega$ with any
significant precision using this technique for a single system.  This
also indicates that the observationally inferred value of $\Omega$
from Virgo infall analysis should be taken with caution.

{\subsection {Dark matter distribution and the Local group}}

The term ``bias" has been coined to relate the global galaxy
distribution to the underlying matter distribution.  Simple physical
arguments (Binney 1977; Rees \& Ostriker 1977; Silk 1977) say that
galaxies should form in regions with special properties: contracting
regions of high density that are relatively ``cool'' should be favored
sites.  Statistical arguments (Kaiser 1984, Bardeen \etal 1986)
indicate that if galaxies form preferentially in density peaks, they
are more strongly clustered than the mass.  Brute-force approaches --
direct hydrodynamic galaxy formation simulations that include both
gravity physics and microphysics on which the above simple arguments
are based, find that galaxies are more strongly clustered than
matter (Cen \& Ostriker 1992; Katz, Hernquist \& Weinberg 1992).

It is interesting to determine if the distribution of galaxies in the
environment surrounding LG binaries is somewhat biased with respect to
the DM one, i.e. if $\delta\rho / \rho$ of halos is larger than
$\delta\rho/\rho$ for the total dark matter distribution.  In this
case it is easier to measure the overdensities in spheres around each
LG. (Fig.13).  We choose a radius of 10 Mpc as representative of the
LG environment, while still being in a mildly nonlinear state.  

We find that halos in the LG environment are slightly biased (b $\sim$
1.5) with respect to the surrounding DM distribution.  This result
holds for both the SCDM and OCDM models, and is consistent with what
is found for a population of bright galaxies in an SCDM Universe
(Kauffmann, Nusser and Steinmetz 1996, Gnedin 1996, Cen \& Ostriker 1992; Katz,
 Hernquist \& Weinberg 1992).
It is remarkable that the same amount of bias is found in the OCDM
cosmology.

\section{Discussion}

We extracted Local Group candidates from large, high resolution 
N-body simulations of two CDM Universes with values of
$\Omega=1.0$ and 0.3. We then compared their properties and the velocity 
field around them with the real Local Group.
To select the binary systems we used a new halo finder algorithm, based
on local density maxima. We show that a binary sample identified with a 
standard ``friend of friends'' algorithm would miss 40$\%$ of them.

The internal properties of the binary groups we
extracted are very similar in both models, even when we relax some of
the constraints on the isolation, separation and radial velocities.
Our result that the Local Group is not a ``special'' place in the nearby
Universe {\it if it was formed in a hierarchical CDM context}, makes our
location extremely useful for testing cosmological models since we can
measure peculiar velocities and distances to many of the nearby
galaxies.

However, we do find large differences in the nearby velocity fields of
these models. These models yield dispersion velocities in the Hubble
flow within a sphere of 5h$^{-1}$ Mpc between 300\---700\kms\ and
150\---300\kms\, respectively. The observed value is 60 \kms.  {\it
Neither of these models can produce a single candidate Local Group
with the observed velocity dispersion in a volume of 10$^6$ Mpc$^3$}.

As far as we know, the Local Group is not special in any way that
could bias these results. We believe that our local overdensity is
fairly well determined, to within a factor of two, and both IRAS and
Optical surveys give $\delta\rho/\rho\sim 0.2$. Even if we lived in a
highly underdense region such that $\delta\rho/\rho \sim -0.3$, our
local Universe could {\it not} be reconciled with a low $\Omega$ CDM
 Universe.

Mixed dark matter models might do better than standard CDM (especially 
SCDM with  a $\sigma_8$ normalization $>0.45$, but they do not perform 
 significantly better
 than the low $\Omega$ model considered here (SDS, their Table 1).
 However, the result of SDS is based on a set of simulations of a smaller
 region of space than ours and the problem of selecting realistic LG's was not
addressed. We expect that a CDM $+{\Lambda}$ model would give a value
of the local velocity dispersions intermediate between SCDM and OCDM.

The distribution of galaxies in the LG environment is biased (b $\sim
1.5$) with respect to the DM distribution, this result is consistent
with other work related to the more general galaxy distribution.

If we ran simulations at a higher resolution, we could resolve more
smaller halos that are in general less clustered. At our present
resolution we are resolving halos that might contain galaxies as
luminous as the Large Magellanic Cloud.  But it is unlikely that, on
average, the bulk motion of very small objects will be significantly lower,
 were they properly resolved and identified, primarily because the 
distributions
of galaxies are observed to be spatially unsegregated in terms of mass
or luminosity.  Therefore, we expect that a higher resolution
simulation would not alter our conclusions significantly.

The inability for a CDM dominated Universe to produce any
Local Group candidates with a cold flow reflects a long standing problem
 of this model if  compared to the observed properties of the local Universe:
 the peculiar velocities of galaxies are too high on average (Gelb \& Bertschinger 1994b)
This evidence may actually suggest that, even when
 normalized to the cluster abundance,  power in CDM--like
 models is too high at intermediate mass scales,
 the more  responsible for peculiar velocities at  scales of a few Mpc.
However, power cannot  at the same time be reduced at small, subgalactic 
scales, as it is necessary  to produce structure at early epochs
 as the observed  Lyman alpha
clouds (Cen \etal 1994; Zhang \etal 1995; Hernquist \etal 1996) or
damped Lyman alpha systems (Mo \& Miralda-Escud\'e 1994; Subramanian
\& Padmanabhan 1994; Kauffmann \& Charlot 1994; Ma \& Bertschinger
1994, Gardner {\it et al} 1996).

Finally,   cosmic variance on scales larger than the volume of our simulations
( $10^6$Mpc$^3$ ) might generate regions where the Hubble flow is substantially
colder than average. Sommerville {\it et al.} (1996) analyzed simulations of the same size
 of ours and found a standard deviation in the pairwise velocity dispersion of
 about 150 \kms for  different observers placed in the same cosmological volume.

Our results also suggest that is critical to obtain new, updated values 
for the peculiar velocities of nearby galaxies, as the problem of the coldness of the Hubble
flow would be  alleviated if the local velocity dispersion had been previously
underestimated. More data should soon be available with the HST Key Project
(Freedman 1994).

The next step for testing the various variants of the CDM model, which are
currently considered to be the most viable family of models will be
large scale galaxy formation simulations. 
It will be necessary to include a  detailed description of 
the physical processes including
hydrodynamics, radiative processes (with possible inclusion of
detailed radiative transfer process) and simple robust star formation
prescriptions (Gelato \& Governato 1996).

A failure of these models to reproduce the basic properties of the nearby
Universe would probably imply that we are missing an important ingredient in
our standard cosmological models. That could be our conception of the
nature of the dark matter itself.

\vskip 0.3truein

\noindent {Acknowledgments}

The simulations were performed at the ARSC and NCSA  supercomputing centers.
This research was funded by the NASA HPCC/ESS program. 
RC would like to thank G. Lake and University of Washington for
the warm hospitality and financial support from the NASA HPCC/ESS Program
during a visit when this work was done.
FG thanks Roger Davies, Sergio Gelato and Luigi Guzzo for useful discussions, and 
the Rolling Stones (hhtp ref:  http://www.stones.com/)  for their continuous
support during the writing of this paper. 

\noindent{\bf References}

\pp Aaronson, M., Huchra, J.P., Mould, J.R., Schechter, P.L., \& Tully, R.B. 1982, ApJ, 258, 64

\pp Bahcall, N.A., \& Cen, R. 1992, ApJL, 398, L81 

\pp Bardeen, J.M., Bond, J.R., Kaiser, N., \& Szalay, A.S. 1986, ApJ, 304, 15.
 
\pp Binney, J.J. 1977, ApJ, 215, 483

\pp Bond, J.R., \& Myers, S.T. 1996, ApJS, 103, 63

\pp Bohringer H., Briel U.G., Schwarz R.A., Voges W, Hartner G. \& Trumper J.
 1994, Nature, 368, 828

\pp Branchini E. \& Carlberg R.G. 1994, {\it Ap.J.}, {\bf 434}, 37.

\pp Brandenberger R.H, Kaplan D.M \& Ramsey S.A.   astro-ph/9310004 

\pp Bunn, E., Scott, D., \& White, M. 1995, ApJL, 441, L9

\pp Bushouse H., Melott A.L., Centrella J. \& Gallagher J.S. 1985,
{\it M.N.R.A.S.}, {\bf 217}, 7p.

\pp Cen R. 1994, {\it Ap.J.}, {\bf 424}, 22.

\pp Cen, R., \& Ostriker, J.P. 1992, ApJ(Lett), 399, L113 

\pp Cen, R., \& Ostriker, J.P. 1993a, ApJ, 417, 404 

\pp Cen, R., \& Ostriker, J.P. 1993b, ApJ, 417, 415

\pp Cen, R., Miralda-Escud\'e, J., Ostriker, J.~P., \& Rauch, M. 1994, ApJ, 437, L9

\pp Dikaiakos, M. \& Stadel, J.  ICS conference proceedings 1996

\pp Dressler, A. 1984, ApJ, 281, 512

\pp Dunn A., M. \& Laflamme R. 1995 {\it Ap.J.}, {\bf 443}, L1.

\pp Eke, V.R., Cole, S., \& Frenk, C.S. 1996,  astro-ph/9601088

\pp Faber, S.M., \& Burstein, D. 1988, in ``Large-Scale Motions
in the Universe: A Vatican Study Week", p115

\pp Fisher K.B., Davis M., Strauss, M.A, Yahil A., Huchra, J. P. 1994 {\it M.N.R.A.S.}, {\bf 267}, 927

\pp Freedman, W.L., 1994, BAAS, 185, 9301 

\pp Gardner J.P, Katz N., Hernquist L. \& Weinberg D.H, 1996 astro-ph/9608142, submitted 

\pp Gelato S., \& Governato F. astro-ph/9610217

\pp Gelb J.M. \& Bertschinger E. 1994a, {\bf 436}, 467

\pp Gelb J.M. \& Bertschinger E. 1994b, {\bf 436}, 491

\pp Giraud E. 1986, {\it Astron.Astrophys.}, {\bf 170}, 1.  

\pp Gnedin, N.Y. 1996, {\it Ap.J.}, {\bf 456}, 1

\pp Governato F., Tozzi P. \& Cavaliere A. 1996, ApJ, 458, 18

\pp Gorski, K.M., Ratra, B., Sugiyama, N., Banday, A.J. 1995, ApJL, 444, L65  

\pp Gunn, J.E. 1978, in ``Observational Cosmology", 8-th Saas-Fee Course, ed.
 by A. Maeder, L. Martinet and G. Tammann (Geneva: Geneva Observatory)

\pp Gunn, J.E., \&  Gott, J.R  1972, ApJ, 176, 1.

\pp Guzzo, L., Fisher, K.B, Strauss, M.A, Giovannelli R. \& Haynes M.P. 1996, {\it preprint}

\pp Hernquist, L., Katz, N., \& Weinberg, D.H. 1996, ApJL, 457, L51

\pp Huchra J. 1988, in ``The Extragalactic Distance Scale", eds. S. van den Bergh \& C.J. Pritchet, Astronomical Society of the Pacific, San Francisco, p257

\pp Hudson M.J. 1993, {\it M.N.R.A.S.}, {\bf 265}, 43

\pp Kahn F.D., Woltjer L., 1959, ApJ, 130, 705

\pp Kaiser, N. 1984, 284, L9

\pp Karachentsev, I. 1996, {\it A.A.}, {\bf 305}, 33.

\pp Katz N., Hernquist, L. \& Weinberg, D.H. 1992, {\it Ap.J}, {\bf 399L}, 109

\pp Kauffmann, G., \& Charlot, S. 1994, ApJ, 430, 97

\pp Kauffmann, G., Nusser, A. \& Steinmetz, M. 1995, astro-ph/9512009 

\pp Kochanek, C.S. 1995, ApJ, 453, 545

\pp Kraan-Korteweg, R.C. 1985, ``The Virgo Cluster", eds. O.-G. Richter \& B. Binggli, ESO, Garching, p200

\pp Kroeker T.L. \& Carlberg R.G. 1991, {\it AP.J.}, {\bf 376 }, 1

\pp Jacoby G.H., Branch D., Clardullo R., Davies R., Harris W.E.,Pierce, M.J.,
 Pritchet C.J., Tonry J.L., Welch D.L., 1992, PASP, 104,599

\pp Lacey, C. \& Cole, C. 1993, {\it M.N.R.A.S.}, 262, 627

\pp Ma, C.-P., \& Bertschinger, E. 1994, ApJL, 434, L5

\pp Marzke, R.O., Geller, M.J., da Costa, L.N. \& Huchra, J.P. 1995, {\it Ap.J} 1995, {\bf 110}, 477.

\pp Mo, H.J., \& Miralda-Escud\'e, J. 1994, ApJL, 430, L25

\pp Moore B. \& Frenk C.S. 1990, in {\it Interactions and Mergers},
ed. R. Wielen, Springer--Verlag Berlin, Heidelberg (1990). pp 410-412

\pp Moore B., Frenk C.S. \& White S.D.M. 1993, {\it M.N.R.A.S.}, {\bf
621}, 827.

\pp Moore B., Katz, N. \& Lake, G. 1996, {\it Ap.J.}, {\bf 457}, 455

\pp Moscardini L., Branchini E., Tini Brunozzi P., Borgani S., Plionis M. 
 \& Coles P. MNRAS, 282, 384

\pp Nusser, A. \& Dekel, A. 1993, {\it Ap.J.}, {\bf 405}, 437.

\pp Ostriker J.P. 1993, ARA\&A, 31, 689 

\pp Oukbir, J., \& Blanchard, A. 1992, A\& A, 262, L21
 
\pp Peebles P.J.E. 1989a, {\it Ap.J.Lett.}, {\bf 344}, L53  

\pp Peebles P.J.E. 1989b, Roy Ast Soc of Canada, 83, 363. 

\pp Peebles P.J.E. 1995, {\it Ap.J.}, {\bf 449}, 52.  

\pp Rees, M., \& Ostriker, J.P. 1977, MNRAS, 179, 451

\pp Tonry, J.L., Ajhar, E.A., Dressler, A., \& Luppino, G.A. 1993, preprint

\pp Sandage A. 1986, {\it Ap.J.}, {\bf 307}, 1.

\pp Schlegel D., Davis M. \& Summers FJ. 1994, {\it Ap.J.}, {\bf 427},
527 (SDS)

\pp Shaya E.J., Peebles P.J.E. \& Tully R.B. 1995, {\it Ap.J.}, {\bf
454}, 15.

\pp Silk, J. 1977, ApJ, 211, 638

\pp Somerville, R.S., Davis, M. \& Primack J.R. 1996, astro-ph/9604041

\pp Somerville, R.S., Primack J.R. \& Nolthenius, R. 1996, astro-ph/9604051

\pp Stadel, J. \& Quinn, T. 1997, {\it in preparation}

\pp Stadel, J., Katz, N., Hernquist, L. \& Weinberg D. {\it in preparation}

\pp Strauss, M., Davis, M., Yahil, A., \& Huchra, J.P. 1992, ApJ, 385, 421 

\pp Subramanian, K., \& Padmanabhan, T. 1994, astro-ph/9402006

\pp Viana, P.T.P, \& Liddle, A.R. 1995, preprint

\pp Villumsen, J., \& Davis, M. 1986, ApJ, 308, 499

\pp White, S.D.M \& Rees, M.J. 1978, {\it MNRAS}, {\bf 183},341

\pp Yahil, A. 1985, ``The Virgo Cluster", eds. O.-G. Richter \& B. Binggli, ESO, Garching, p359

\pp Yahil, A., Sandage, A., \& Tammann, G.A. 1980, ApJ, 242, 448

\pp Zhang, Y., Anninos, P., \& Norman, M.L. 1995, ApJL, 453, L57

\vskip 0.5truein

\noindent{\bf Figure captions}

\noindent{\bf Figure 1} The upper panel shows the density of the mass
distribution of the SCDM simulation in a $5\times 50 \times 50$
h$^{-1}$ Mpc slice, higher densities are brighter colors.  The lower
panel shows the distribution of halos found by Skid, the colors are
random.

\noindent {\bf Figure 2} A ``Local Group'' binary system found by Skid.
The lines connected to the particle points show the trajectories
towards the local density maxima. Note the  small satellites
between the two main galaxies.

\noindent{\bf Figure 3} A histogram of halo circular velocities of all the LG halos,
 plotted
as a fraction of the total number.  The dot-dashed red line are halos
belonging to  LG candidates  in  the SCDM simulation (a) 
The green dashed line are from
the OCDM simulation (b) . The solid line in each panel  shows 
the distribution of halos from the general samples from each simulation.  
Panel (c) compares the mass functions of SCDM and OCDM.

\noindent{\bf Figure 4} A histogram of relative radial velocities
 of the binary halos within
the control sample of binaries (solid white lines) and  Local
Groups (dot-dashed red line are SCDM groups (a), dashed green line are
OCDM groups (b)).  Panel (c) compares the SCDM and OCDM distributions.

\noindent{\bf Figure 5} The relative radial velocities of the binary halos
 are plotted against
their separation.  Panel (a) are OCDM systems, green dots (stars in
the b/w plot) are LG binaries and black dots are from the general
sample.  Panel (b) are SCDM systems, red open circles are LG binaries
andblack dots are from the general sample.

\noindent {\bf Figure 6} The relative radial velocities of the binary
 halos divided by the
total relative velocity are plotted as a histogram for the OCDM model
(a) and SCDM model (b). The colors are as in Figure 3. Panel (c)
compares the two models.

\noindent{\bf Figure 7} The density field in
 a $10\times 50 \times 50$ h$^{-1}$ Mpc slice is
plotted for the  OCDM simulation.  Green
boxes show the location of Local Group candidates in each model. The
color scale  has been stretched such that the plotted
densities span a range between the average and $10^4$ times the
average value for each model. Since the models have the same phases in
the initial conditions, with this scaling the structures appear very
similar. However, the SCDM simulation has more massive halos.

\noindent{\bf Figure 8} Same for the SCDM model.

\noindent{\bf Figure 9} A model Local Group from the SCDM simulation 
that is located in a planar structure. This Group is featured in Movie
2.

\noindent {\bf Figure 10} A histogram of the overdensities of the
 LG environments (dot-dashed
red line are SCDM groups, green dashed line are OCDM groups).  The
density is calculated on a sphere of 10 Mpc around each LG.

\noindent {\bf Figure 11} The dispersion of peculiar velocities from the
 local Hubble flow for
each Local Group. Overdensity is measured in a sphere of 5h$^{-1}$Mpc.
Red open circles are SCDM groups and green filled circles are OCDM groups.
The observed value is plotted as a star.

\noindent {\bf Figure 12}  The estimated value of $\Omega$ plotted against
 a convenient function of
 overdensity of the cluster-group region. 
$\Omega$ is calculated using the ``Virgo-centric infall model'' 
applied to two candidates from (a) the OCDM simulation  (green full dots)
and (b) for the  six candidates in the SCDM simulation  (red circles).

\noindent {\bf Figure 13} A plot of the bias in the halo distribution
 of each Local Group
environment (red open circles are SCDM halos, green dots are OCDM
halos).  The line shows the relation that would be found when there
is no bias.

\noindent {\bf Movie 1} This MPEG movie slices through the OCDM cube. 
Particles belonging
to LGS  are in green. Bright colors represent higher density regions.
LGs hinabit a wide variety of environments, , but are preferentially located
in filamentary or planary structures. The slice thickness is 5Mpc.

\noindent {\bf Movie 2} This MPEG movie shows a region of space centered on 
a LG extracted from the  SCDM
volume. The rotation shows clearly the planar structure in which the LG 
(LG particles are marked in green) is embedded. 
The complex network of filaments that forms the whole surrounding region is 
evident. The same filaments continue well outside the spherical region of space
shown in the animation. The sphere diameter is 20Mpc.

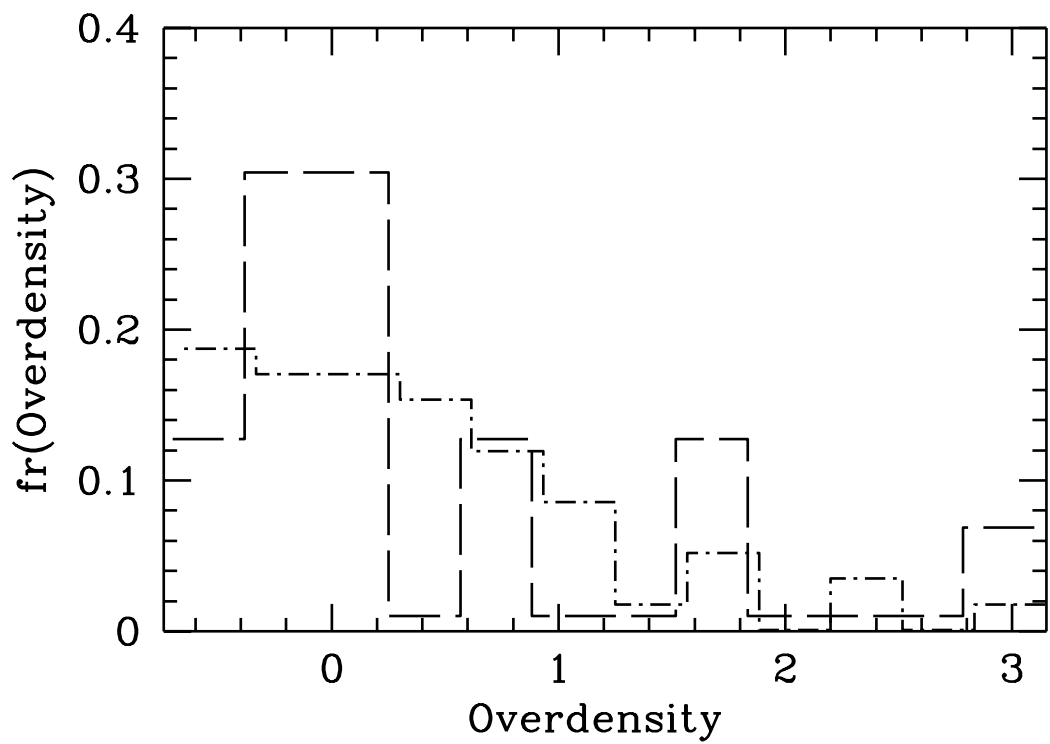
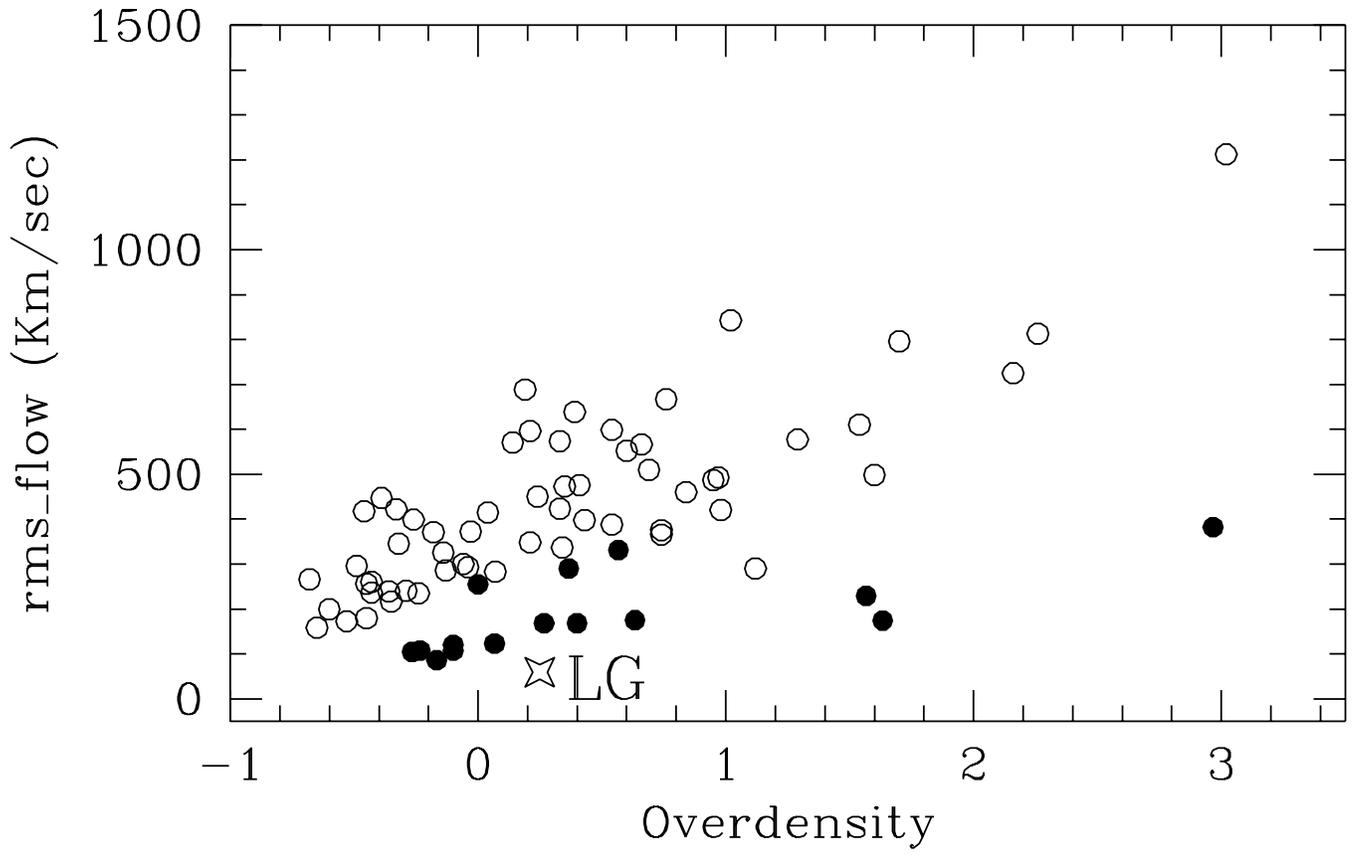
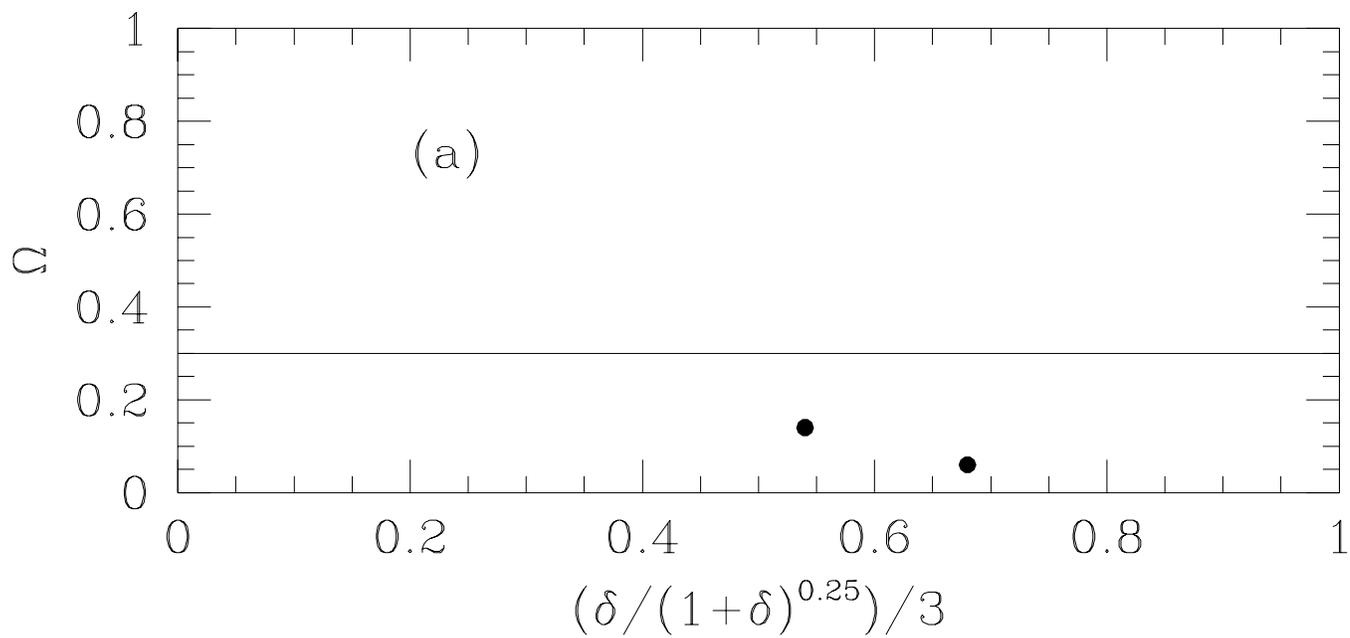
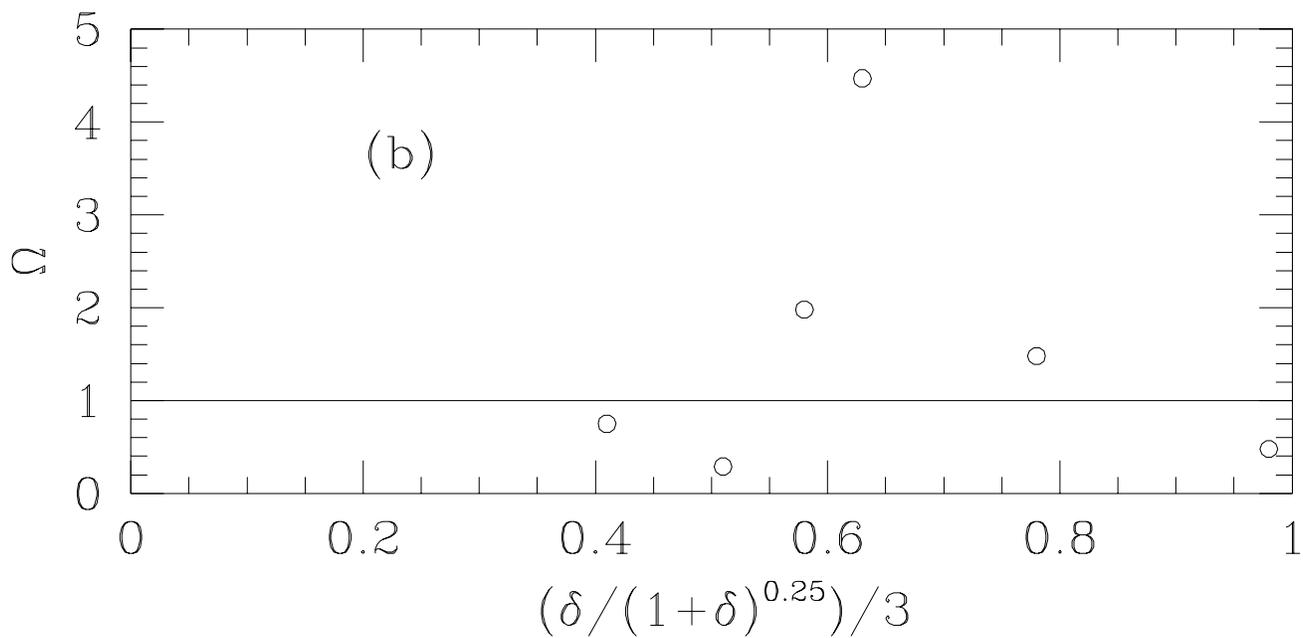
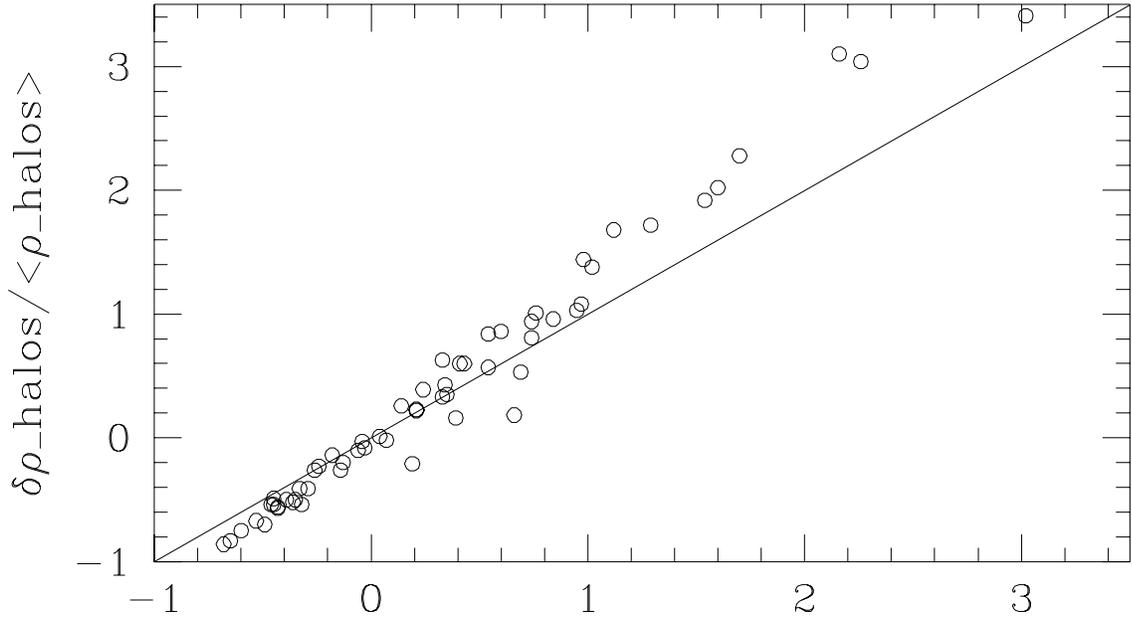
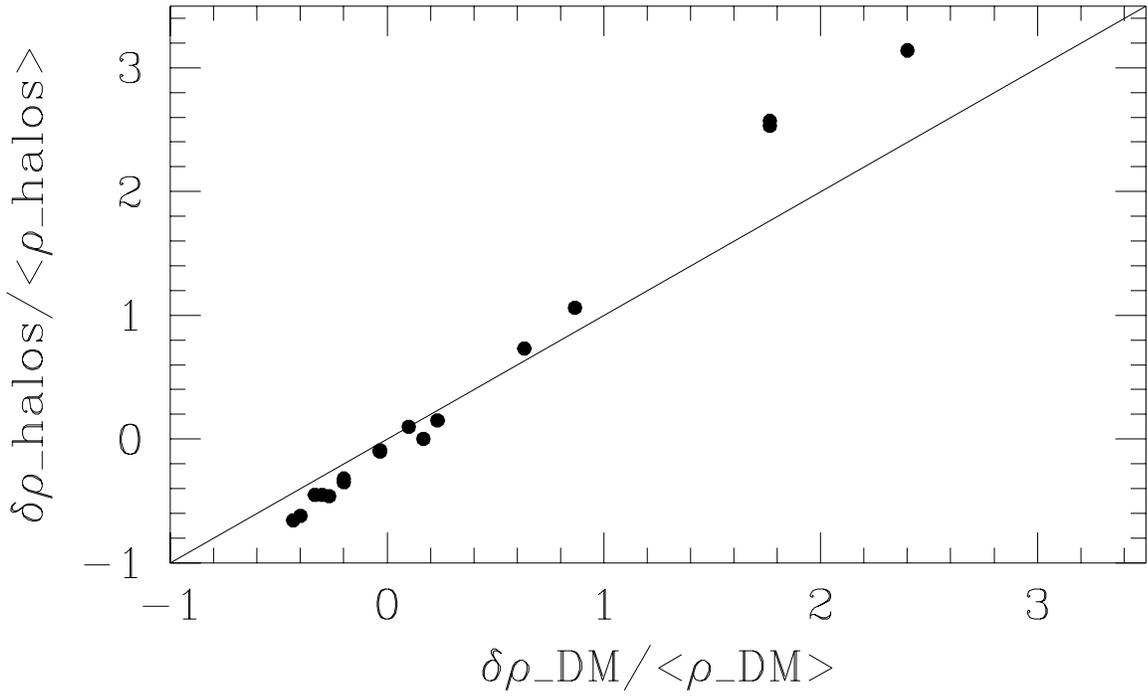
\end{document}